\documentclass{aastex63}

\usepackage{mathtools}
\usepackage{CJK}
\usepackage{subfigure}
\usepackage{graphicx}
\usepackage{color}
\shorttitle{The energy transport of sloshing oscillation in coronal loop}
\shortauthors{Xia et al.}

\begin{document}
\begin{CJK*}{UTF8}{gbsn}
\title{Plasma heating and nanoflare caused by slow-mode wave in a coronal loop}

\correspondingauthor{Yang Su}
\email{yang.su@pmo.ac.cn}
\correspondingauthor{Tongjiang Wang}
\email{wangt@cua.edu}

\author[0000-0002-2630-4753]{Fanxiaoyu Xia}
\affil{Key Laboratory of Dark Matter and Space Astronomy, Purple Mountain Observatory, Chinese Academy of Sciences, Nanjing 210033, China}
\affil{School of Astronomy and Space Science, University of Science and Technology of China, Hefei 230026, China}
 
\author[0000-0003-0053-1146]{Tongjiang Wang}
\affil{The Catholic University of America at NASA Goddard Space Flight Center, Solar Physics Laboratory, Code 671, Greenbelt, MD 20771, USA}

\author[0000-0002-4241-9921]{Yang Su}
\affil{Key Laboratory of Dark Matter and Space Astronomy, Purple Mountain Observatory, Chinese Academy of Sciences, Nanjing 210033, China}
\affil{School of Astronomy and Space Science, University of Science and Technology of China, Hefei 230026, China}

\author[0000-0003-3160-4379]{Jie Zhao}
\affil{Key Laboratory of Dark Matter and Space Astronomy, Purple Mountain Observatory, Chinese Academy of Sciences, Nanjing 210033, China}

\author[0000-0003-4078-2265]{Qingmin Zhang }
\affil{Key Laboratory of Dark Matter and Space Astronomy, Purple Mountain Observatory, Chinese Academy of Sciences, Nanjing 210033, China}

\author{Astrid M. Veronig}

\affil{Kanzelh\"ohe Observatory for Solar and Environmental Research, University of Graz, Universit\"atsplatz 5A-8010 Graz, Austria}

\author{Weiqun Gan}
\affil{Key Laboratory of Dark Matter and Space Astronomy, Purple Mountain Observatory, Chinese Academy of Sciences, Nanjing 210033, China}

\begin{abstract}
We present a detailed analysis of a reflecting intensity perturbation in a large coronal loop that appeared as sloshing oscillation and lasted for at least one and a half periods. The perturbation is initiated by a microflare at one footpoint of the loop, propagates along the loop and is eventually  reflected at the remote footpoint where significant brightenings are observed in all the AIA extreme-ultraviolet (EUV) channels. This unique observation provides us with the opportunity to better understand not only the thermal properties and damping mechanisms of the sloshing oscillation, but also the energy transfer at the remote footpoint. Based on differential emission measures (DEM) analysis and the technique of coronal seismology, we find that 1) the calculated local sound speed is consistent with the observed propagation speed of the perturbation during the oscillation, which is suggestive of a slow magnetoacoustic wave; 2) thermal conduction is the major damping mechanism of the wave but additional damping mechanism such as anomalous enhancement of compressive viscosity or wave leakage is also required to account for the rapid decay of the observed waves; 3) the wave produced a nanoflare at the remote footpoint, with a peak thermal energy of $\thicksim10^{24}-10^{25}$ erg. This work provides a consistent picture of the magnetoacoustic wave propagation and reflection in a coronal loop, and reports the first solid evidence of a wave-induced nanoflare. The results reveal new clues for further simulation studies and may help solving the coronal heating problem. 
\end{abstract}

\section{Introduction}

Standing slow magnetoacoustic wave in hot (T $\ge$ 6 MK) coronal loops are usually called "SUMER" oscillations. They were confirmed for the first time by the Doppler shift oscillations in observations of the \textit{Solar and Heliospheric Observatory}/Solar Ultraviolet Measurement of Emitted Radiation (\textit{SOHO}/SUMER) instrument \citep{2002ApJ...574L.101W}. The phase speed of the oscillation estimated using the observed loop length and the oscillation period was found to be close to the sound speed in the loop, which is consistent with the characteristics of slow-mode waves. By verifying a quarter-period phase shift between the intensity and the velocity oscillations, further evidence for the slow-mode wave nature of the SUMER oscillations was provided by \cite{2003A&A...406.1105W}. \cite{2011SSRv..158..397W} reviewed the observational and theoretical studies of SUMER oscillations, and suggested that small (micro) flares at the loop footpoints are most likely the trigger of these events. It was also pointed out that strong damping and quick excitation of the standing slow-mode waves in flaring coronal loops are an important characteristics that desire further studies, since the knowledge of these physical processes can be used to diagnose the poorly-known energy transports in coronal structures by using a technique called coronal seismology \citep{2005LRSP....2....3N,2020ARA&A..58..441N}.

Recently, propagating reflected slow-mode waves that were predicted to exist in coronal loops \citep[see the review by][] {wan21} have attracted a lot of attention. Propagating slow-mode waves in a hot coronal loop (also called sloshing oscillation) were first reported by \cite{2013ApJ...779L...7K} in observations of the \textit{Solar Dynamics Observatory}/Atmospheric Imaging Assembly (\textit{SDO}/AIA). They were identified as a propagating and reflecting intensity perturbation along the loop as observed in AIA 94 {\rm \AA} and 131 {\rm \AA} channels, with a propagation speed close to the sound speed of the plasma in the loop (like in the SUMER oscillations). \cite{2016ApJ...828...72M} analyzed these oscillations in hot loops as slow-mode waves using coordinated \textit{Hinode}/X-ray Telescope (XRT) X-ray and \textit{SDO}/AIA EUV observations for the first time. Based on a 2.5D MHD simulation \citep{2015ApJ...813...33F} and using the parameters derived from the observations as input, they also suggested that the main damping mechanism is thermal conduction. In contrast to this case, \cite{2015ApJ...811L..13W} found observational evidence for compressive viscosity to be the major damping mechanism while thermal conduction is strongly suppressed in a different case. By simulating the oscillation event in \cite{2015ApJ...811L..13W} with a 1D nonlinear MHD model, \cite{2018ApJ...860..107W} and \cite{wang19} found that the modeled wave features better match the observation when the observationally-constrained transport coefficients are used instead of the classical values. They also predicted that the propagating mode can quickly transform into a standing mode within one oscillation period when compressive viscosity is anomalously enhanced. 

In a statistical study, \cite{2019ApJ...874L...1N} showed that both the sloshing and standing oscillations in coronal loops are most likely slow magnetoacoustic waves, whereas some SUMER oscillations may initially be propagating waves. They presented a detailed analysis of the relationships between the damping time and other parameters for both the SUMER and the sloshing oscillations, and suggested that the competition between the dissipative processes and nonlinearity effect may account for the observed decay of the sloshing oscillations. Transformation from an impulsively excited propagating wave into a standing mode with the help of compressive viscosity \citep{2018ApJ...860..107W} or leakage of wave energy through the footpoints to lower layers \citep[e.g.,][]{2005A&A...436..701S} or also laterally to the corona \citep[e.g.,][]{2007A&A...467..311O,2009A&A...495..313O,Ofman_2022} was predicted by MHD simulations. \cite{2021ApJ...914...81K} have recently found observational evidence to support this prediction, they have also suggested that thermal conduction is the major damping mechanism in the event they studied. Both observations and simulations in previous studies suggest that the reflected propagating slow wave can be triggered by the reconnection process at the footpoint of the loop. Some recent studies showed observational evidence for the reconnection in the fan-spine magnetic topology \citep[e.g.,][]{2015ApJ...804....4K,2018ApJ...860..107W}. However, the mechanisms for the damping and excitation of the slow-mode oscillations in hot loops are still inconclusive.

In this paper, we report an unique event of sloshing oscillations in a coronal loop detected by \textit{SDO}/AIA. {We find a remote brightening (RB) occurring nearly simultaneously in all EUV channels during the reflection of the waves. We propose that the RB in this event could be caused by the energy loss of slow waves mainly due to thermal conduction at the footpoint.} The observations and temperature obtained from differential emission measures (DEM) analysis are described in Section \ref{sec:sec2}. The thermal properties of the wave are derived in Section \ref{sec:sec3} and the damping mechanisms are discussed in Section \ref{sec:sec4}. The energy transport between the wave and the remote brightening is analyzed in section \ref{sec:sec5}. The discussion and conclusions are given in Section \ref{sec:sec6}.

\section{Overview}\label{sec:sec2}

\begin{figure*}[ht!]
	\begin{minipage}{1\textwidth}
		\centering
		\includegraphics[height=1.\textwidth,angle=90.]{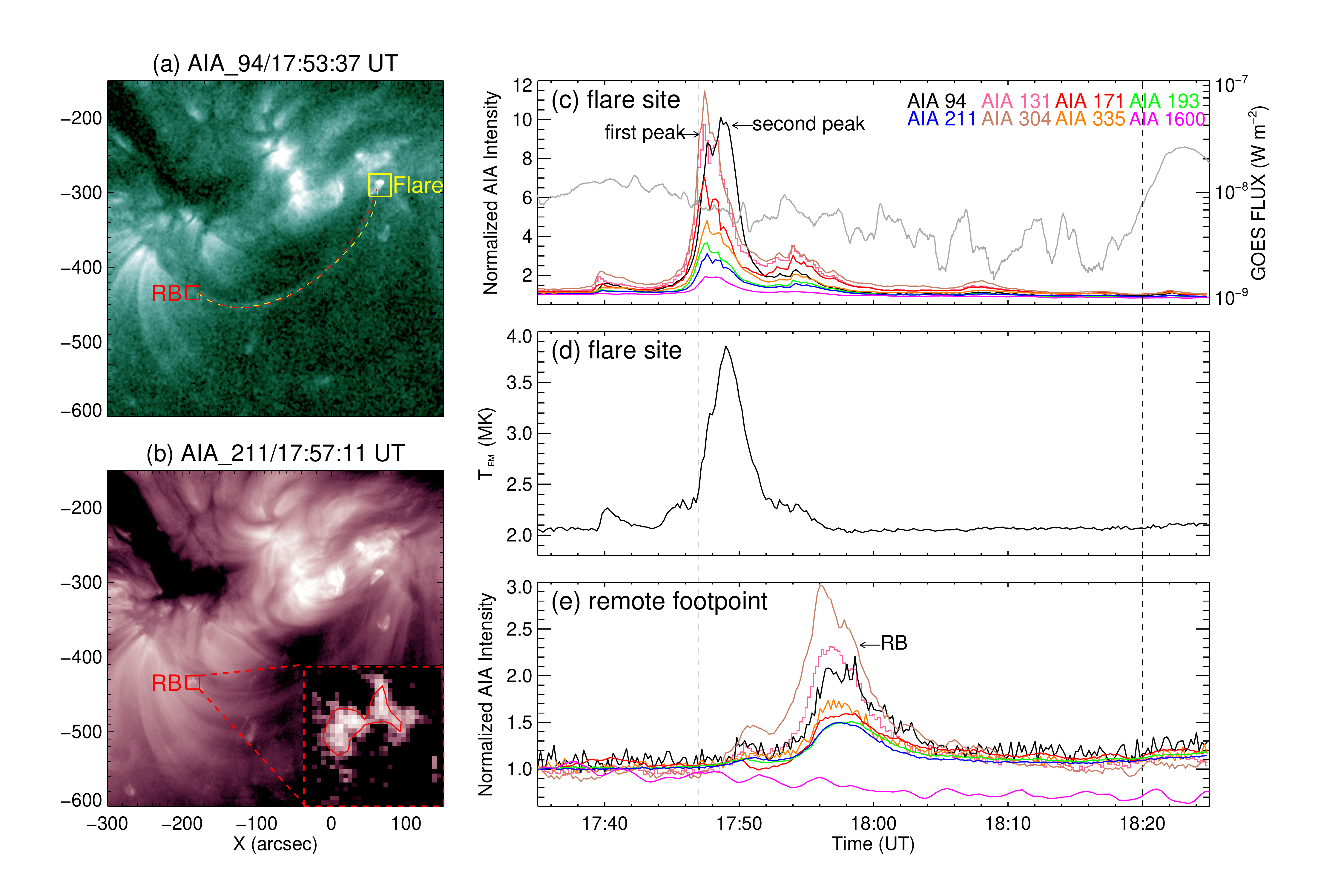}
		\caption{\scriptsize{Overview of the observations from \textit{SDO}/AIA and the temperature derived from DEM analysis. (a) AIA 94 {\rm \AA} image of the loop region. The yellow and red dashed curves show the manually traced loop and the fitted loop, respectively. The yellow and red boxes mark the regions of the flare and the remote brightening, respectively. (b) AIA 211 {\rm \AA} image of the loop region. Zoom-in at right bottom corner shows the base difference image of the region of interest. The red contour corresponds to the 80$\%$ level of the maximum value of the AIA 211 {\rm \AA} image. (c) Light curves derived by summing over the yellow box for the different AIA filters (94 {\rm \AA}, 131 {\rm \AA}, 171 {\rm \AA}, 193 {\rm \AA}, 211 {\rm \AA}, 304 {\rm \AA}, 335 {\rm \AA}, and 1600 {\rm \AA}) normalized to the pre-event values at 17:37 UT. GOES 1.0-8.0 {\rm \AA} light curve is shown in gray. (d) Evolution of the plasma temperature averaged over the flare region. (e) AIA light curves derived by summing over the red box. The dashed vertical lines in panels (c)-(e) indicate the start and end times of the oscillation.} \label{figure1}}
	\end{minipage}
\end{figure*}

The Atmospheric Imaging Assembly \citep[AIA,][]{2012SoPh..275...17L} onboard the Solar Dynamics Observatory \citep[\textit{SDO},][]{2012SoPh..275....3P} provides images of the Sun in 10 extreme ultraviolet (EUV) and UV-visible channels with a temporal cadence as high as 12 seconds. In our study, a microflare is observed by AIA on 2014 June 10 near the edge of NOAA Active Region (AR) 12085, with the start time at $\sim$17:45 UT and end time at $\sim$17:51 UT. During the impulsive phase, an emission enhancement is detected in the AIA 94 {\rm \AA} channel at the flare site. It propagates to the east side along a large coronal loop and then reflects back. The enhancement propagates between the remote footpoint and the flare site and lasts for at least one and a half periods before fading away. At $\sim$17:54 UT, when the reflection happens, a weak EUV brightening occurs at the remote footpoint of the loop (marked in Fig.~\ref{figure1}(b)), located in a quiet region.

Using the 94 {\rm \AA} images during the oscillation (Fig.~\ref{figure1}(a)), we manually track the propagation to trace the loop (marked by the yellow dashed curve) connecting the region of the flare (yellow box) and the remote footpoint (red box). To calculate the loop length more accurately, we fit the derived 2D curve using a 3D semicircular loop model \citep{2002SoPh..206...69S,2002ApJ...574L.101W,2009SSRv..149...31A}. The 2D projection of the fitted loop is shown by the red dashed curve in Fig.~\ref{figure1}(a). The best-fit parameters obtained for the loop geometry are the following: the loop center offset from the solar surface $h_{0}  = -0.09R_{\odot}$ (solar radii), the inclination angle from the vertical $\theta = 30.7^{\circ}$, and the loop length $L \thickapprox$ 270 Mm. 
 

Fig.~\ref{figure1}(c) displays the intensity evolution of intensities summed over the flare region in different AIA channels. The microflare is not recorded in the GOES flare list but definitely, it is below GOES A-class, according to the 1-8 {\rm \AA} soft X-ray profile. While the flare is detected by the AIA 1600 {\rm \AA} channel and all the EUV channels (Fig.~\ref{figure1}(c)), the intensity oscillation is detected only in the AIA 94 {\rm \AA} filter (dominated by Fe \uppercase\expandafter{\romannumeral18}, formed at T $\thickapprox$ 6.3 MK), which implies that the perturbation is hot. We notice that a second, higher intensity peak shows up only in the 94 {\rm \AA} light curve at $\sim$17:49 UT, which may be related to the propagating intensity disturbance triggered at about the peak time of the flare ($\sim$17:47 UT). 

Using the latest version of the revised Sparse DEM code \citep[]{2015ApJ...807..143C,2018ApJ...856L..17S} and binning the AIA images merged by $2 \times 2$ pixels, we calculate the DEMs for each binned pixel individually over the logarithmic temperature range of [5.5, 7.6] K. The EM-weighted average temperature $T_{EM}$ calculated for the flare region is shown in Fig.~\ref{figure1}(d). We find that the temperature of the plasma in this flare region peaks roughly simultaneously with the second peak in the 94 {\rm \AA} time profile.


Time profiles of AIA intensities summed over the remote footpoint region (Fig.~\ref{figure1}(e)) showed that, the remote brightening (RB) is visible in all the EUV channels at the time when the propagating perturbation arrives ($\sim$17:54 UT). However, no enhancement in the AIA 1600 {\AA} channel is observed. These features may imply that the RB is caused by the propagating perturbation, during its reflection at the coronal base before it can reach the lower atmosphere (the chromosphere and the transition region).
  
\section{Thermal Properties of the Wave}\label{sec:sec3}

\begin{figure*}[ht!]
	\begin{minipage}{1\textwidth}
		\centering
		\includegraphics[height=1.\textwidth,angle=90.]{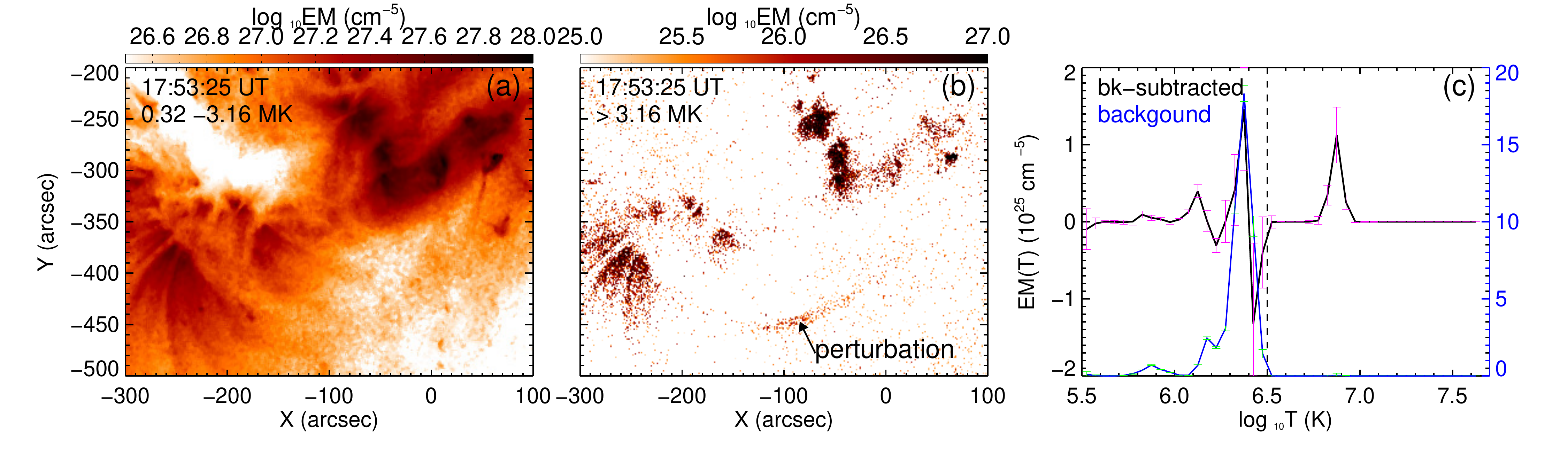}
		\caption{\scriptsize{DEM analysis for the loop region. (a) and (b) DEM maps in different temperature ranges. (c) The background subtracted EM($T$) profile (black) and the background EM($T$) profile (blue) for the perturbation region marked in panel (b). The error bars derived from the Monte Carlo method provided by the revised sparse DEM code are shown in pink and green. {The vertical dashed line indicates the lower boundary to separate the spike-shaped hot component above log$T$ = 6.5 K from the warm background.}} \label{figure2}}
	\end{minipage}
\end{figure*}

To measure the thermal properties of the propagating perturbation, we calculate the DEMs in the whole loop region at 17:53 UT (within the first oscillation period), and integrate the EM($T$) in temperature intervals of [0.32, 3.16] MK and $>$ 3.16 MK, respectively. The EM images (Fig.~\ref{figure2}(a) and (b)) clearly show the presence of the propagating perturbation in the hotter plasma. The DEMs before the oscillation are used as an estimated of the background plasma. This is further confirmed by the EM($T$) profile of the heated plasma (Fig.~\ref{figure2}(c)) in the perturbation region (marked by an arrow in panel (b)), which manifests as a single spike well separated from the warm coronal background. The error bars are obtained from 100 realizations using the Monte Carlo method. We measure the DEM peak at $T_{h} \sim$7.0 MK and its FWHM width $w \sim$1.6 MK. The cooler component in the selected region is associated with the emission from the coronal background plasma and does not change significantly.

\begin{figure}[ht!]
	\begin{minipage}{1\textwidth}
		\centering
		\includegraphics[height=0.9\textwidth,angle=90.]{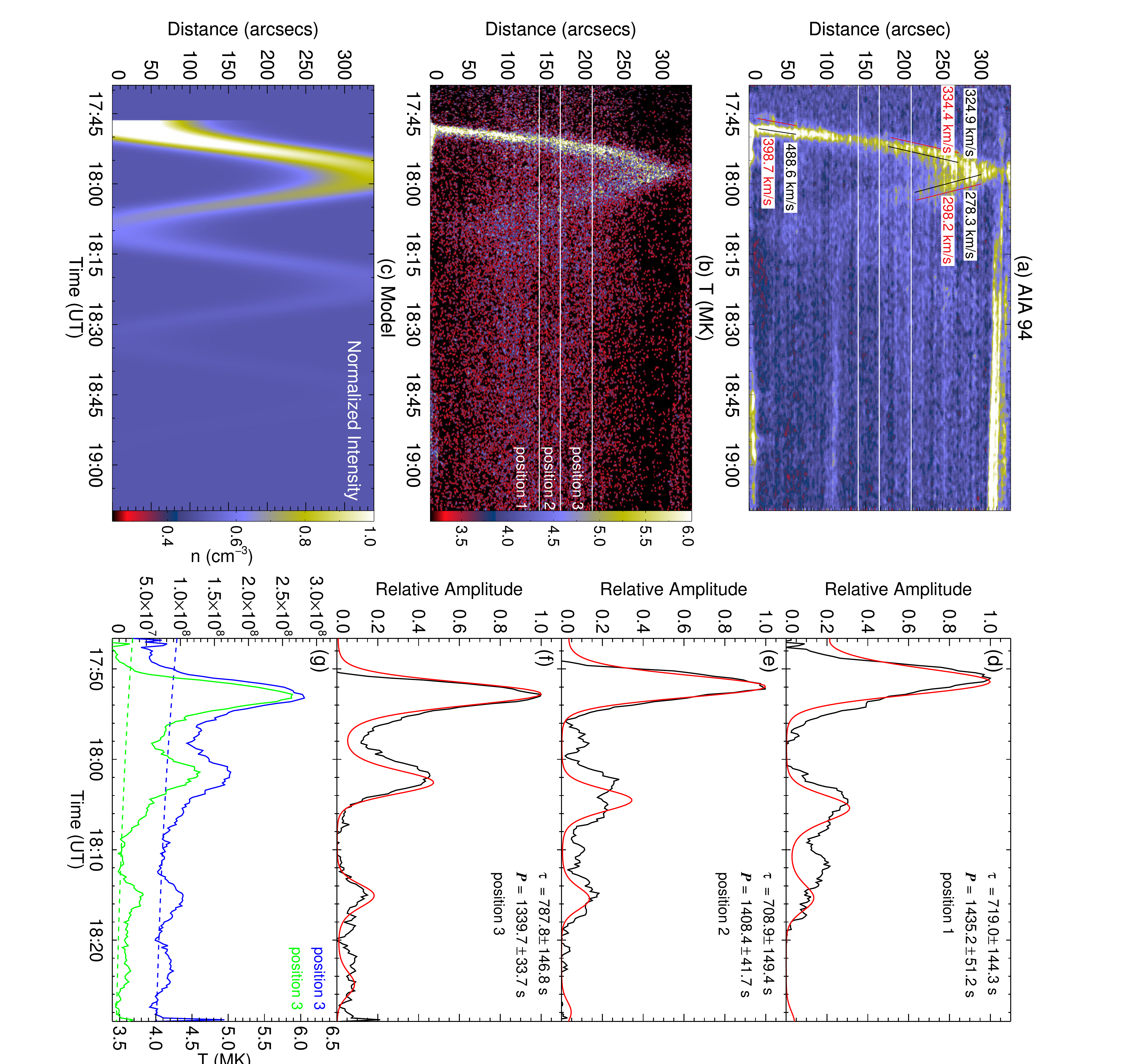}
		\caption{\scriptsize{(a) Time-distance map of the intensity from the base-difference images along the traced loop in Fig.~\ref{figure1}(a). The red lines indicate the sound speeds calculated from the mean temperatures, and the fitted propagating speeds are shown in black. (b) Time-distance map of the temperature along the loop. The horizontal lines in panel (a) and (b) mark the selected positions 1$-$3 along the loop. (c) Time-distance map of intensity modelled with the best-fit parameters of the oscillation. (d)-(f) The detrended and normalized light curves (black) along with the best-fit light curves (red) for position 1$-$3. (g) The electron density (blue) and temperature (green) for the oscillation (solid line) and background trend (dashed line) at position 3.} \label{figure3}}
	\end{minipage}
\end{figure}

To study the physical characteristics of the propagating intensity perturbation, we extract the base-difference intensities along the loop slice (marked by the yellow dashed curve in Fig.~\ref{figure1}(a)) from the AIA 94 {\rm \AA} images. The base image is taken at 17:38:49 UT and the intensities along the slice are averaged over a width of 7 pixels (4.2{\arcsec}). To emphasize the propagation of the perturbation, we detrend and normalize the base-difference intensity at each spatial location along the loop by the mean background intensities after the oscillation. The derived time-distance map (Fig.~\ref{figure3}(a)) over the time range 17:39$-$19:10 UT clearly shows that the intensity perturbation initiates from the flare site, propagates along the loop, then is reflected at the remote footpoint, lasting at least one and a half periods before fading. The detailed evolution can be seen in the time profiles (Fig.~\ref{figure3}(d,e,f), see also Sec. \ref{sec:sec6}) for three selected positions (indicated in Fig.~\ref{figure3}(a,b)). The characteristics of the reflection suggest that the intensity perturbation is a propagating slow-mode wave (pulse), in agreement with theoretical predictions based on 1D MHD simulations \citep[see][]{2008A&A...481..247T,2016ApJ...826L..20R,2018ApJ...860..107W}. 

In the next step, we select three different time intervals from the time-distance map and fit the bright ridges to the propagation speeds (marked in black in Fig.~\ref{figure3}(a)). The local sound speed of the plasma, through which the wave passed during the three time intervals, is estimated using the DEM results. To this aim, we calculate the EM-weighted mean temperature solely from the hot (log$T$ \textgreater 6.5 K, see the vertical line in Fig.~\ref{figure2}(c)) component in each pixel, which provides a better representation of the perturbation than considering the EMs over the entire temperature range. The time-distance plot of the obtained temperature along the loop (Fig.~\ref{figure3}(b)) shows that the temperature disturbance is reflected back and forth almost simultaneously with the intensity disturbance, and its amplitude decreases with the wave propagation. We estimate the perturbed temperature to be $T$= 6.88, 4.84 and 3.85 MK during the three time intervals. By $C_{s} \sim152\sqrt{(T/\rm {MK}}$) km~s$^{-1}$, for the local adiabatic sound speeds, we obtain $C_s=398.7$, 334.4, and 298.2 km~s$^{-1}$ (the values are also marked in red in Fig.~\ref{figure3}(a)), which are found in good agreement with the measured propagation speeds from the linear fitting. The results clearly indicate that the intensity disturbance is a slow magnetoacoustic wave rather than a flowing plasma blob \citep[e.g.,][]{Su_2012}.


\section{Damping Mechanisms}\label{sec:sec4}
\subsection{Measurements of physical parameters}
\begin{figure}[ht!]
		\centering
		\includegraphics[width=0.95\textwidth]{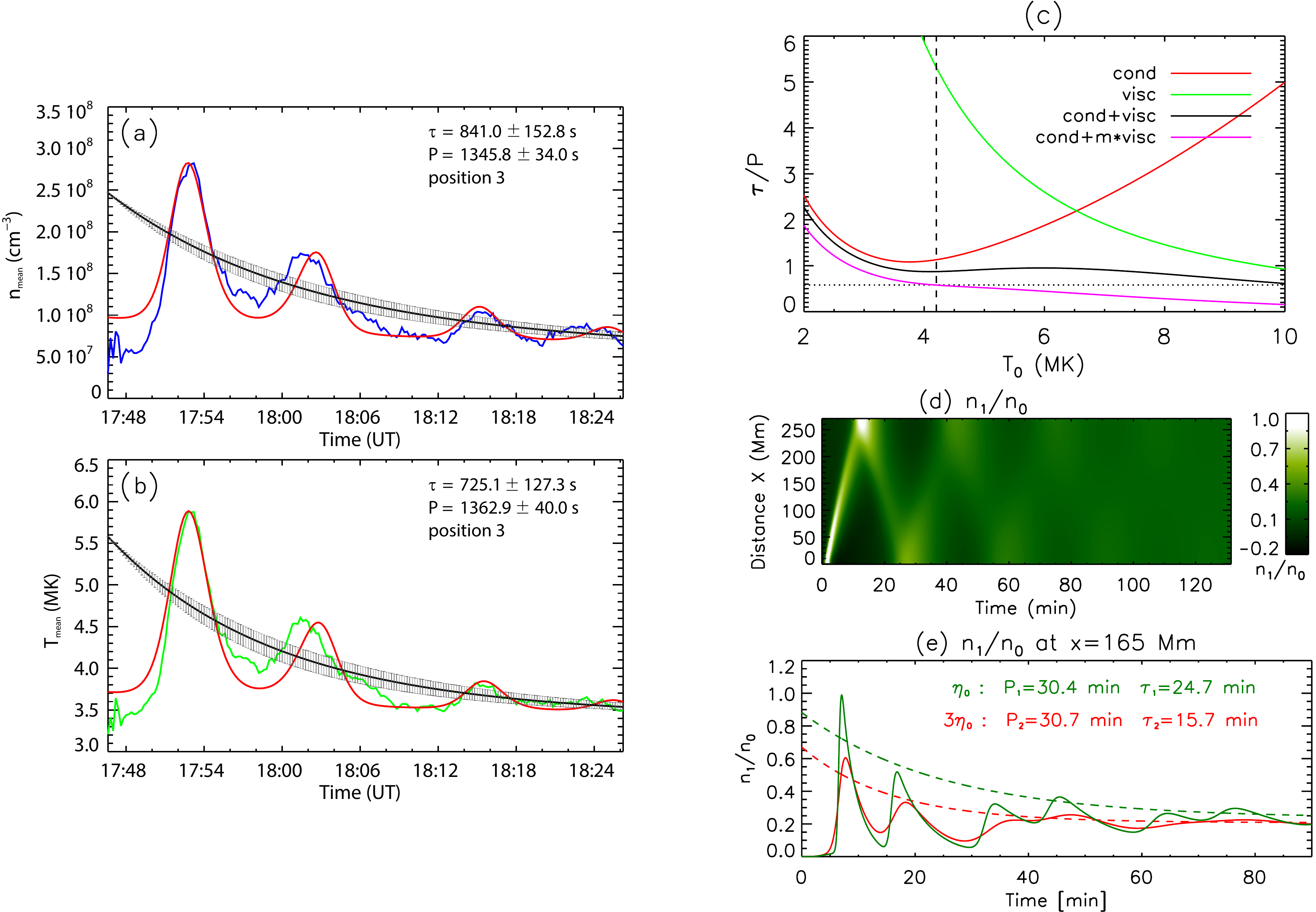}
	\caption{\scriptsize{(a) The evolution of plasma density $n$ and temperature $T$ derived from the observations at position 3 at the loop are shown in blue and green, the fitted light curves of $\Delta n$ and $\Delta T$ plus the background trends are shown in red, and the mean of plasma density $n_{\rm mean}$ and temperature $T_{\rm mean}$ are shown in black. The error bars of $n_{\rm mean}$ and $T_{\rm mean}$ are shown in gray. (c) The ratio of the damping time ($\tau$) to the wave period ($P$) as a function of the loop equilibrium temperature calculated based on the linear wave theory with thermal conduction alone (red), compressive viscosity alone (green), considering both (black), and the case with the viscosity enhanced by a factor of 3 (pink). The vertical dashed line marks the mean temperature ($T_0$=4.2 MK) of the observed loop, and the horizontal dotted line marks the observed $\tau/P$=0.58. (d) Time distance map of the perturbed density ($n_1/n_0$) along the loop simulated using a 1D nonlinear MHD model with the classical thermal conduction and the 3-times enhanced viscosity (see the text in detail). (e) Time profiles of the perturbed density at the location $x$=165 Mm for the one case when both the thermal conduction and viscosity coefficients take the classical values (green) and for the other case with the classical thermal conduction but the 3-times enhanced viscosity (red). The dashed lines are their exponential decay time fits.The measured wave period and damping time are marked on the plot.} \label{figure4}}
\end{figure}

Recently, \cite{2021ApJ...914...81K} pointed out that the exponentially-damped sine function cannot accurately fit the light curves extracted from various locations for the sloshing oscillation, since the time profiles of the perturbation are dependent on both the spatial location along the loop and the spatial width of the perturbation. They proposed a toy model to fit the sloshing oscillations, considering the spatial location and the width of the perturbation:
\begin{equation}
I (x, t) = A_{0} e^{-\frac{t}{\tau}} e^{-\frac{{(x-x_{0})}^2}{{\sigma}^2}} \label{equation1}
\end{equation}
where $A_{0}$ is the amplitude, $t$ is time, $\tau$ is the decay time, $x$ is the spatial location of the extracted light curve, $x_{0} = \frac{L}{2} (1 - \cos(\frac{2\pi t}{P}))$ is the spatial location, and $\sigma$ is the spatial width of the perturbation. $L$ and $P$ are the length of the loop and the period of the oscillation, respectively. We fit the time evolution of the mean intensities (averaged over 8.7 Mm along the loop) obtained from positions 1$-$3 on the loop (shown by the horizontal lines from bottom to top in Fig.~\ref{figure3}(a) and (b)) using this model, after the original light curve at each position was detrended and normalized using the background generated by the post-oscillation intensity. The detrended and normalized light curves for positions 1$-$3 are displayed in panels (d)-(f) of Fig.~\ref{figure3}.

In our study, the obtained light curves are evidently different from sinusoidal curves, because the perturbation has a short spatial width (compared to the loop length) and short duration. In order to reduce projection effects, the selected positions are not close to the footpoints, and the length of the selected regions is short. The loop length $L$ and the distance $x$ are determined by the 3D loop geometry. We fit the time profiles to Equation \ref{equation1} using \textbf{MPFIT} with the free parameters $\tau$, $P$ and $\sigma$. The best-fit curves and results are shown in panels (d)-(f) of Fig.~\ref{figure3}, the mean of the best-fit parameters are $\tau$ = 740.0 $\pm$ 146.8 s, $P$ = 1394.5 $\pm$ 42.2 s and $\sigma=58\pm17$ Mm. Using these results, we reconstruct a time-distance map of intensity oscillation using Equation \ref{equation1} with the measured parameters superposed on a constant background. Fig.~\ref{figure3}(c) shows that the toy model is very close to the observed one. 

To calculate the density and temperature of the plasma in the loop, we derive the density $n = \sqrt{{\rm EM}/w}$ and temperature $T$ at positions 1 and 3 from the observations, and fit the profiles of the background-subtracted plasma density $\Delta n = n - n_{\rm bg}$ and temperature $\Delta T = T - T_{\rm bg}$ to Equation \ref{equation1}. $EM$ and $T$ are obtained from EM($T$) in temperature intervals of [3.16, 39.81] MK according to the DEM results at positions 1 and 3. The loop width $w$ is assumed to be equal to the line of sight (LOS) depth, which is estimated to be about 4.8{\arcsec} from a Gaussian fit to the intensity profile across the loop. As shown in Fig.~\ref{figure3}(g), the background trends of the density and temperature ($n_{\rm bg}$ and $T_{\rm bg}$) are obtained by fitting their background values to an exponential decay function. The fitted $n$ and $T$ for position 3 are shown in Fig.~\ref{figure4}(a) and (b). Then, we obtain the mean density and temperature of plasma at position 1 and 3 by $n_{\rm mean} = {\rm mean}(\Delta n) + n_{\rm bg}$ and $T_{\rm mean} = {\rm mean}(\Delta T) + T_{\rm bg}$ (as shown in Fig.~\ref{figure4}(a) and (b)), where ${\rm mean}(\Delta n)$ and ${\rm mean}(\Delta T)$ are the half amplitudes of $\Delta n$ and $\Delta T$. By considering the density $n_{0}$ and temperature $T_{0}$ of plasma in equilibrium for the whole loop as the mean values of $n_{\rm mean}$ and $T_{\rm mean}$ for position 1 and 3 during the first period, we obtain $n_{0} = (1.5\pm0.2)\times10^8$ cm$^{-3}$ and $ T_{0} = (4.2\pm0.3)$ MK. The plasma density is smaller than the typical loop density in ARs due to the fact that the large loop is not located in a registered AR and that only plasma with log$T$ [K] \textgreater 6.5 is considered here.


\subsection{Interpretations}
In the following, we investigate the damping mechanism for the observed slow waves by considering non-ideal MHD effects including thermal conduction, compressive viscosity, and optically thin radiation based on linear theory, and then test the results using a 1D nonlinear MHD modeling. We first consider radiative losses. For the observed loop with $T_0$=4.2 MK, $n_0=1.5\times10^8$ cm$^{-3}$, and length $L$=270 Mm, we estimate the radiation cooling timescale to be $\sim1.29 \times 10^{5}$ s using the formula given in \citet[][]{2013ApJ...778..139S, 2015ApJ...811L..13W}:
\begin{equation}
\tau_{\rm rad} = 3450\frac{T_{6}^{3/2}}{n_{9}},  
\label{equation2}
\end{equation}
where $T_6 = T [{\rm K}]/10^6$, $n_{9} = n [{\rm cm}^{-3}]/10^{9}$. The obtained $\tau_{\rm rad}$ is at least two orders of magnitudes larger than the oscillation period. Therefore, the effect of radiative losses on damping is negligible. This result is consistent with previous studies reporting that the radiative damping may become important only in a cooler and dense loop \citep[e.g.,][]{pan06,prasad21,wan21}, which is, however, not the case here.

The effect of thermal conduction on wave dissipation can be quantified by the thermal
ratio $d$ \citep[e.g.,][]{dem03}, defined as
\begin{equation}
 d=\frac{1}{\gamma}\frac{P_0}{\tau_{\rm cond}}=3.75\times{10^4}\left(\frac{T_0^2}{n_0L}\right),
 \end{equation}
where $P_0=2L/C_s$, $\tau_{\rm cond}$ is the thermal conduction timescale and $\gamma = 5/3$ is the adiabatic index. It is known that the ratio $\tau/P$ reaches a minimum value of 1.1 at $d\approx$0.1 \citep[e.g.,][]{wan21}. For the hot loop analyzed here we have $d$=0.16, so the ratio $\tau/P$ is expected to be slightly larger than its minimum. Indeed, solving the dispersion relation for thermal conduction, we obtain $(\tau/P)_{\rm cond}$=1.13 (see the red curve in Fig.~\ref{figure4}c). This implies that the dissipation by thermal conduction alone is insufficient to explain the quick decay of the observed waves with $(\tau/P)_{\rm obs}$=0.58.

From the dispersion relation for compressive viscosity alone \citep[e.g.,][]{sig07, wang19}, we obtain 
\begin{equation}
     \frac{\tau}{P} \approx \frac{3}{8\pi^2}\left(\frac{1}{\epsilon}\right),
\end{equation}
where $\epsilon$ is the viscous ratio, defined as
\begin{equation}
 \epsilon=\frac{\eta_0}{\rho_0 C_s^2 P_0}=1.65\times{10^3}\left(\frac{T_0^2}{n_0L}\right),
\end{equation}
where $\eta_0=10^{-16}T_0^{5/2}$ is the classical Braginskii compressive viscosity coefficient \citep[see][]{holl86}, and $P_0=2L/C_s$. With the measured loop parameters, we obtain $\epsilon=7.2\times10^{-3}$ and $(\tau/P)_{\rm visc}$=5.3 (see the green curve in Fig.~\ref{figure4}c). This indicates that the damping by compressive viscosity is about 5 times weaker than that by thermal conduction in the case under study.

To estimate the combined effect of thermal conduction and compressive viscosity, we numerically solve the dispersion relation including both effects using Equ.~15 in \cite{mac10}. For a loop with an equilibrium density $n_0=1.5\times10^{8}$ cm$^{-3}$, we derive the dependence of $\tau/P$ on the equilibrium temperature as shown in Fig.~\ref{figure4}c (black solid curve). For a loop at $T_0$=4.2 MK, we obtain $P$=2069 s and $\tau$=1804 s, giving a ratio  $(\tau/P)_{\rm cond+visc}$=0.87 for the fundamental mode. This implies that the damping due to the combined effect of thermal conduction and viscosity is still weaker by about 30\% than the observed damping. Motivated by the study of \citet{2015ApJ...811L..13W} who found evidence for a significant enhancement of compressive viscosity from the observed damping of slow-mode waves by coronal seismology, we modify the viscosity coefficient while keeping the thermal conduction coefficient as the classical value. By matching the predicted value of $\tau/P$ to the observed one, using an optimization method (see the pink curve in Fig.~\ref{figure4}c), we find that the viscosity coefficient needs to be enhanced by a factor of $m$=3.1. Note that damping by thermal conduction still dominates over viscous damping even if the viscosity is enhanced by a factor of 3, because the ratio of $(\tau/P)_{\rm visc}^{\rm en}$=1.7 for the enhanced viscosity alone is larger than the value of $(\tau/P)_{\rm cond}$ for thermal conduction alone.

On the other hand, the observed oscillations show a large amplitude which is more than 50\% relative to the mean density ($n_{\rm mean}$) for the first peak (see Fig.~\ref{figure4}a). When compared to the background trend ($n_{\rm bg}$), the relative amplitude is even larger (up to several times of $n_{\rm bg}$; see Fig.~\ref{figure3}g). Large amplitudes can cause important nonlinear effects and produce enhanced damping due to shock dissipation \citep[e.g.,][]{verw08,rud13}. Using 1D MHD simulations, we examine the role of nonlinear effects on wave damping in the presence of thermal conduction and compressive viscosity to verify the results obtained from the linear theory above. We excite the slow magnetoacoustic waves by injecting a flow pulse from one end of an initially uniform coronal loop. The loop model is set up with the observed parameters $T_0$, $n_0$, and $L$, and the flow pulse is taken as the same form as used in \citet{2018ApJ...860..107W,wang19}. We take the flow amplitude $V_0=C_s$=340 km~s$^{-1}$ for $T_0$=4.2 MK and the pulse duration $t_{\rm dur}$=180 s. We simulate the wave excitation for two cases: one with thermal conduction and compressive viscosity coefficients both taken as the classical values, and the other with the classical thermal conductivity but the viscosity coefficient enhanced by a factor of $m$=3. Fig.~\ref{figure4}d shows the temporal evolution of the perturbed densities ($n_1/n_0=(n-n_0)/n_0$) along the loop for the second case. This simulation can approximately reproduce the observed features: the excited propagating wave is reflected back and forth for about three times, and then tends to transition into a standing mode. However, the observed signals are too weak to confirm the prediction for the formation of a standing mode in the late phase. In Fig.~\ref{figure4}e we compare the time profiles of density perturbations at a location $x$=165 Mm (similar to position 3 in the observed case) between the two simulated cases. We obtain a wave period ${P_1}$=30.4 minutes and damping time $\tau_1$=24.7 minutes, respectively for the first case, while $P_2$= 30.7 minutes and $\tau_2$= 15.7 minutes for the second case. We find that the ratio $\tau_2/P_2$=0.51 for the enhanced viscosity case matches the observation ($(\tau/P)_{\rm obs}$=0.58) much better than the classical viscosity case corresponding to $\tau_1/P_1$=0.81. This result also indicates that the enhanced viscosity effectively suppressed the role of nonlinearity in the wave damping when the amplitudes are large.


\section{Energy Transport between the Wave and Remote Brightening}\label{sec:sec5}

\begin{figure*}[ht!]
	\begin{minipage}{1\textwidth}
     	\centering
		\includegraphics[width=1.\textwidth]{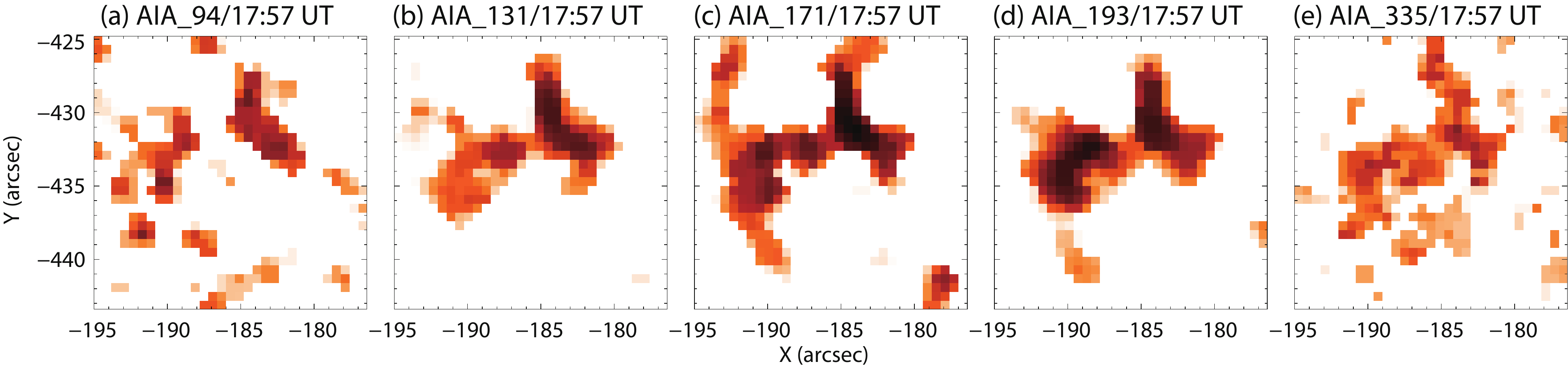}
		\centering
		\includegraphics[height=0.9\textwidth,angle=90.]{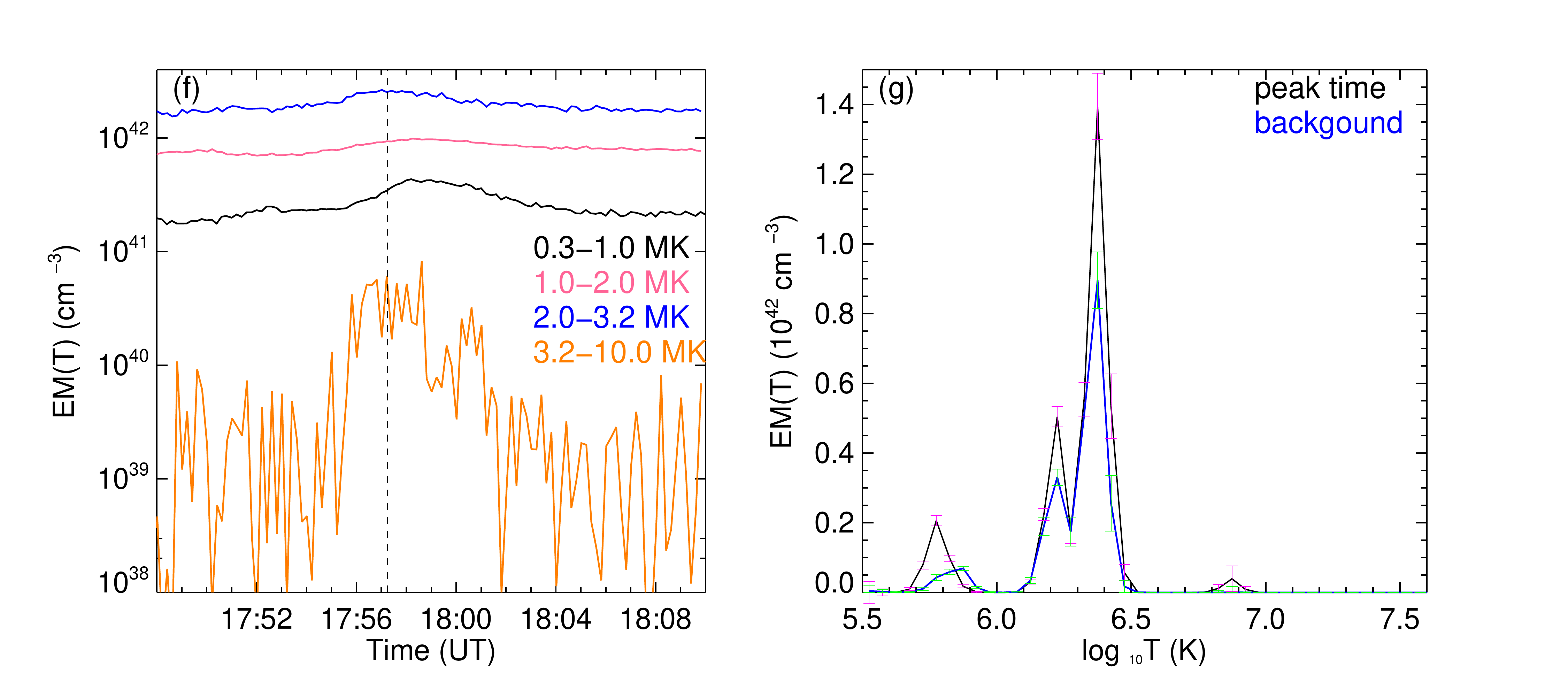}
		\caption{\scriptsize{(a)-(e) Base difference images for the region of the remote brightening (RB). (f) Time profiles for EM($T$) integrated over different temperature intervals. The peak time of the thermal energy is marked by the vertical line. (g) For the region of the RB, EM($T$) profiles at the peak time of thermal energy (black) and the time before RB (blue) derived with the Monte Carlo method.} \label{figure5}}
	\end{minipage}
\end{figure*}

The energy flux of the slow magnetoacoutsic wave is given by $F = 0.5\cdot\rho v^{2}C_{s}$ \citep[e.g.,][]{1999ApJ...514..441O}, where $v$ is the velocity amplitude of the waves, which can be estimated from $v \sim {\Delta n}/n\cdot C_{s}$ based on linear theory, and $\rho$ is the mass density. We estimate the power of the slow magnetoacoutsic wave at a selected position $x$ on the loop by $P(x,t) = 0.6 m_{p} \cdot {\Delta n}'(x,t)^2/n_{\rm mean}(x,t) \cdot {C_{s}(x,t)}^3\cdot A$, where ${m_{p}}$ is the proton mass, $A$ is the area of the loop cross section, and the sound speed $C_{s}(x,t)$ is calculated using $T_{\rm mean}(x,t)$. $n_{\rm mean}(x,t)$ and $T_{\rm mean}(x,t)$ are the mean number density and the mean temperature of the plasma, respectively, and ${\Delta n}'(x,t) = {\Delta n}(x,t) - {\rm mean}({\Delta n}(x,t))$ is the amplitude of the density perturbation at position $x$. 

The initial energy carried by the wave can be estimated by assuming that the initial power of the wave is a constant during the pulse. By fitting the pulse which induces the wave (the second peak in Fig.~\ref{figure1}(c)) with a Gaussian function, the duration of the pulse is obtained as the FWHM of the Gaussian profile, giving $\approx$165 s. Using the fit results of ${\Delta n}$ and ${\Delta T}$ for the position near the footpoint, the peak power of the wave at the footpoint is about 5.25 $\times10^{23}$ erg~s$^{-1}$, which is assumed to be the initial power of the wave. In this way, the total energy of the slow wave is estimated to be $\sim 8.7 \times 10^{25}$ erg.
 
The brightenings at the remote footpoint caused by the propagating wave are visible in all the AIA EUV channels. We further investigate the brightenings with DEM analysis. Clear enhancements of EM are found in the time profiles which are summed over the brightened area and four different temperature ranges (Fig.~\ref{figure5}(f)). More detailed comparison of the plasma EM($T$) during the peak time (marked by a vertical dashed line in Fig.~\ref{figure5}(f)) with its background shows an increase in EM at different temperatures in the range 0.3$-$10 MK and the appearance of a hot component at $\sim$ 7 MK (Fig.~\ref{figure5}(g)), which is a clear manifestation of a heating process.

In order to study the energy transport between the wave and the RB, we analyze two different aspects. On one hand, the total energy released by the wave at the remote footpoint (e.g., due to non-ideal effects including thermal conduction, viscous heating, and radiative cooling) is estimated. To this aim, we first derive the powers of the wave reaching and leaving position 3 using the method mentioned above, which are $1.75\times10^{23}$ and $0.5\times10^{23}$ erg~s$^{-1}$, respectively. Then by assuming that the power of the wave pulse keeps constant during its passage, we obtain an upper limit of the energy budget for the RB to be $\sim 2.1\times10^{25}$ erg from the decrement in wave energy at position 3. On the other hand, we estimate the total thermal energy of the RB plasma from the EM($T$) maps by $E_{th} = \sum_{k}3k_{B}T_{k}\sqrt{{\rm EM_{k}}V_{RB}}$, where $k$ is the index of temperature bins $T_k$ in 0.3$-$10 MK, $k_{B}$ is the Boltzmann constant, ${\rm EM_{k}}$ [cm$^{-3}$] is the total EM integrated over the area of the RB at temperature $T_{k}$, and $V_{RB} = A_{RB} \cdot D_{RB}$ is the volume of the RB. $A_{RB}$ is selected to be the area of the contour at a level of 80$\%$ of the peak intensity in the AIA 211 {\rm \AA} image (Fig.~\ref{figure1}(b)). The LOS depth $D_{RB}$ of the heated footpoint is determined in two ways: one is the widely used estimate of $D_{RB} = \sqrt{A_{RB}}$, and the other one is the distance from the lower corona to the top of the chromosphere (roughly the formation depth for AIA 1600 {\AA}), i.e., $\sim$2{\arcsec} from \cite{Vernazza1981ApJS}. The latter one might be a better estimate for the present event. From these two approaches, we obtain the background-subtracted peak thermal energy of the RB to be $\sim 1.12\times10^{25}$ erg and $\sim 0.77\times10^{25}$ erg, respectively, both lower than the released energy of the wave at the remote footpoint. These results further support the interpretation that the RB could result from the heating by dissipation of the slow wave at the remote footpoint.

In order to better understand the relation between the microflare and the reflected slow wave in the loop, we also estimate the peak thermal energy of the microflare which gives 6.8 $\times 10^{26}$ erg. This is almost 8 times as large as the initial wave energy. The findings suggest that the microflare can provide the required energy for the excitation of the observed slow wave, and that the trapped wave energy corresponds to about 13\% of the total energy released in this microflare.

\section{Discussion and Conclusions} \label{sec:sec6}

\subsection{The slow-mode wave} 
We present a comprehensive study of a longitudinal oscillation in the AIA 94 {\rm \AA} channel, which is a manifestation of a propagating and reflected intensity perturbation in a large coronal loop. For the first time, we analyze the evolution of the thermal properties of the sloshing oscillation with time and distance, which are crucial for understanding the damping mechanism. 

The consistency between the sound speeds derived from the DEM analysis and the propagation speeds of the observed intensity perturbation for different time intervals during the oscillation reveals the nature of the perturbation as slow magnetoacoustic wave. The time profiles of the intensity oscillations at different locations along the loop are fitted with the model introduced by \cite{2021ApJ...914...81K}, to measure the wave parameters more accurately. We obtain the period, decay time, and perturbation width to be $P$ = 1394.5 $\pm$ 42.2 s and $\tau$ = 740.0 $\pm$ 146.8 s, and $\sigma=58\pm17$ Mm, respectively. 

While the spike-shaped DEM profile of the heated plasma implies that the loop may have a transverse structure in density and temperature, the damping due to the transverse structure is expected to be unimportant. Using the measured DEM peak of $T_{h}\sim$7 MK and FWHM width of $W$=1.6 MK, and assuming the temperature contrast between the interior and exterior of the loop as $T_{i}/T_{e}=(T_{h}+W/2)/(T_{h}-W/2)$, we obtain $T_{i}/T_{e}$=1.3 and $C_{i}/C_{e}$=1.1. The standard cylinder model predicts that slow mode waves in the corona are only weakly dispersive (propagating with the tube speed $C_{t}$ meeting $C_{e}<C_{t}<C_{i}$) and are trapped for all wave numbers of the longitudinal waves \citep[e.g.,][]{1984ApJ...279..857R}. This implies that in the ideal MHD case, a slow-mode pulse will nearly maintain its shape during the propagation. Theoretical and numerical studies have shown that lateral leakage of slow waves into the ambient corona (e.g., by oblique propagation and model coupling) is generally insignificant in the low-$\beta$ corona \citep[e.g.,][]{2015A&A...573A..32A}. Therefore, this mechanism alone cannot explain the rapid damping rate (with $\tau/P\sim$1) typically observed in hot flaring loops \citep[e.g.,][]{2009A&A...495..313O}.

As suggested by previous numerical simulations of slow waves, the damping due to wave leakage at footpoints \citep[e.g.,][]{2005A&A...436..701S,2005A&A...438..713T} or in the corona \citep[e.g.,][]{2007A&A...467..311O} is generally inefficient compared to the major damping mechanism such as thermal conduction. Especially for long wavelength components such as the fundamental mode, it takes most of the total wave energy based on the Fourier power spectral analysis and so dominates in wave amplitudes \citep{2018ApJ...860..107W}. Using 1D MHD simulations, \cite{2005A&A...436..701S} were the first to study the effect of energy leakage at footpoints on wave damping, which was found to be small with $\tau/P\ga$5 for both the fundamental and second harmonics. Later on, including a realistic stratification of the solar atmosphere (chromosphere, transition region, and corona) in the 1D MHD model,  a number of studies have been carried out to contribute to the understanding of the excitation and damping mechanisms for standing waves \citep{2005A&A...438..713T,2007ApJ...659L.173T,2008A&A...481..247T,2008A&A...483..301B} and sloshing oscillations \citep{2016ApJ...826L..20R,2018ApJ...856...51R,2019ApJ...884..131R}. It was shown that thermal conduction is the dominant damping mechanism in typical hot flaring loops. The damping caused by wave leakage into the chromosphere is too weak to account for the observed rapid damping \citep{2005A&A...438..713T}. A similar conclusion was obtained by a 2.5D MHD model with a realistic magnetic geometry and stratification \citep{2015ApJ...813...33F}. In addition, the effect of radiative cooling on the damping is negligible for hot and under-dense coronal loops \citep[e.g.,][]{pan06,prasad21,wan21}. Therefore, we analyze the damping mechanisms by solving the dispersion relation only including the wave dissipation by thermal conduction and compressive viscosity. With the measured physical parameters of the loop, we find that in a large and hot coronal loop, thermal conduction is the dominant damping mechanism but additional damping by compressive viscosity with an enhancement 3 times as large as the classical value is required to account for the quick decay as observed. Because the observed waves show large amplitudes of more than 50$\%$ of the mean density, we verified our result above using a 1D nonlinear MHD model constrained by the observations. We find that the enhanced compressive viscosity greatly reduces the nonlinear effect for waves of large amplitude, thus our conclusion obtained from linear theory remains valid. In addition, our 1D model with the modified transport coefficients successfully reproduces the observed sloshing oscillation by injecting a strong and short flow pulse from one footpoint of the loop.

However, the effects of wave leakage cannot be neglected in some conditions. Based on 3D MHD simulations without considering the effect of thermal conduction, \cite{Ofman_2022} found that wave leakage due to mode coupling with kink mode may affect the damping of slow waves more efficiently than compressive viscosity (with the classical value) in a hot coronal loop of 2-6 MK with realistic magnetic geometry. However, their assumption of neglecting thermal conduction is not suitable here because thermal conduction is found to be a main damping mechanism based on the linear theory and MHD simulations in our study. In addition, our observations do not show any signatures for the presence of kink oscillations coupled with the sloshing oscillations, and thus it does not support a strong coronal leakage in this case. \citet{2007ApJ...668L..83S} and \citet{2009A&A...495..313O} quantitatively studied the role of lateral wave leakage through a nonuniform boundary layer due to wave refraction in curved loops using a 2D arcade model. The ideal MHD simulations by \cite{2009A&A...495..313O} showed $\tau/P$=1.85 due to wave leakage in a curved slab. If assuming that such a damping rate by wave leakage is linearly overlaid on that by thermal conduction and viscosity, we can estimate their combined effect from $1/(\tau/P)_{\rm comb}=1/(\tau/P)_{{\rm cond}+{\rm visc}} + 1/(\tau/P)_{\rm leak}$ and obtain $(\tau/P)_{\rm comb}$=0.59, very close to the observation. {Thus, wave leakage to the remote footpoint and to the corona may provide a possible mechanism required for the additional damping in our study, which may eventually make it unnecessary to invoke an anomalous enhancement in compressive viscosity.} In addition, since the high harmonics are more liable to be leaked outside through the nonuniform transverse structure of the loop, this may explain the suppression of shock development despite the observations of waves with large amplitudes \citep{2015A&A...573A..32A}. To confirm this scenario, advanced 3D MHD simulations including a more realistic AR model and flare heating function are required in the future.

The estimated total energy carried by the slow wave is $\sim$8.7 $\times10^{25}$ erg, which comprises $\sim13$\% of the peak thermal energy of the microflare that triggered the sloshing oscillation. The finding supports that the sloshing oscillation can be excited by a microflare. This was also found in previous studies \citep[e.g.,][]{2016ApJ...828...72M}. 


\subsection{The nanoflare at the remote footpoint} 

Thermal energy of remote EUV brightenings can be transported from a flare by various means like energetic particles, shock waves, thermal conduction, or plasma flows \citep{1988ApJ...326..451M, 2013A&A...557L...5Z, 2017ApJ...847..124H}. In this work we find that slow magnetoacoustic waves excited by flares can also transport energy to a remote site and produce EUV brightenings.


While the slow wave is only observed in the AIA 94 {\AA} channel, the brightening caused by the slow wave at the remote footpoint is visible in all the EUV channels. The reconstructed DEM profiles show an enhancement that occurs nearly simultaneously at different temperatures from 0.5 to 8 MK. This feature is clearly distinct from heating due to adiabatic compression in the wave front, where the EM decreases in the lower temperature channel (e.g., seen as a dimming in the 171 \AA\ channel), while the EM increases in the higher temperature channels (e.g., seen as brightenings in the 193 and 211 \AA\ channels for large-scale EUV waves \citep{vann15, liu12}.

The energy estimate from our DEM analysis indicates that heating in the remote brightening regions is significant. The peak thermal energy of the RB is estimated to be $\sim10^{25}$ erg, about $12\%$ of the initial wave energy. Indeed, the total kinetic energy of the waves $E_k=\int_0^L\rho(x)V(x)^2{\rm d} x$ at the time ($t$=3.3 min) when the flow injection is stopped, and the times ($t$=7.5 and 18.4 min for the first two peaks in Fig.~\ref{figure4}e) before and after the reflection can be calculated using our simulation results from the model with 3$\eta_0$ (see Fig.~\ref{figure4}d). We find that 29\% of the initial wave energy is lost due to the dissipation by thermal conduction and compressive viscosity. This prediction is consistent with our estimate of the energy loss (24\% of initial wave energy) from the observations using the variations of $n$ and $T$, {thus providing a strong support to our scenario as discussed below. We suggest that the loss of wave energy due to thermal conduction that transfers heat from the hotter footpoint region into the chromosphere or the transition region during the reflection accounts for the RB, because the most of wave energy is expected to be in the form of internal energy during the reflection (as the footpoints are natural velocity nodes). This process may be regarded as “leakage of wave energy" out of a coronal loop system through its low boundary, since the strong heat flux flowing from the wave-induced hotter footpoint region down to the chromosphere due to thermal conduction will shift the transition region downward \citep[see numerical simulations by][]{2015ApJ...813...33F}. It needs to be pointed out that the RB may also be produced by energetic particles or thermal front originating from the flare region \citep[e.g., ][]{2013ApJ...778..139S}, however, such processes would predict the occurrence of the RB much earlier than the arrival of slow waves, in disagreement with the observation in this event.} The thermal energy of the RB is significantly lower than the energy released in a microflare \cite[$10^{26} - 10^{28}$ erg, e.g., ][]{2007SoPh..246..339S,Hannah2011SSRv,Cooper2020ApJ} but comparable to the energy budget for a nanoflare proposed by \cite{1988ApJ...330..474P} and that in other following works \cite[e.g.,][]{Testa2014Sci,Bahauddin2021NatAs}. Our result provides the first evidence of a nanoflare that is directly produced by a slow magnetoacoustic wave. 

Wave heating and nanoflare reconnection heating are so far the two most promising mechanisms \cite[][and the references therein]{Moriyasu2004ApJ,Antolin2008ApJ} for solving the coronal heating problem. It has been pointed out that the dissipation of Alfv\'{e}n waves can produce nanoflares \cite[e.g., ][]{Moriyasu2004ApJ,Antolin2008ApJ}. Our work adds a link between slow waves and nanoflares by revealing that slow waves can not only heat the plasma in coronal loops but also heat the plasma in the lower atmosphere at footpoints to temperatures in a wide range from 0.5 MK to 8 MK, producing a nanoflare. However, whether this kind of heating events generally exist and how they play a role in the coronal heating need further investigation. 
 
 

 

\acknowledgments
\textit{SDO} is a mission for NASA's Living with a Star program. This work is supported by the National Natural Science Foundation of China (grant Nos. 11820101002, U1631242, 11921003, U1731241, 41774183, 41861134033, U1931138) and the Strategic Pioneer Program on Space Science, Chinese Academy of Sciences (grant Nos. XDA15320300, XDA15016800, XDA15320104, XDA15052200). T.W. is grateful to supports by NASA grants 80NSSC18K1131, 80NSSC18K0668, 80NSSC21K1687, 80NSSC22K0755 as well as the NASA Cooperative Agreement
80NSSC21M0180 to CUA. A.V. acknowledges the Austrian Science Fund (FWF): I 4555-N. F.X. processed the data, analyzed the results, performed the calculations, compared observations with models, made four of the figures, and wrote the draft manuscript. T.W. preformed the simulation and theoretical calculations, helped with important interpretations of the data and results. Y.S. found the case, initiated the study, prepared the EUV data, analyzed the DEMs, provided initial result and methodology. All authors discussed on the interpretation of the results, the method, and the remote brightenings, helped improving the figures and the paper.

\vspace{5mm}

\bibliographystyle{aasjournal}
\bibliography{ref}
\end{CJK*}
\end{document}